\documentclass[12pt]{article}
\usepackage{texdraw}
\usepackage{latexsym,amsfonts,amsmath,amsthm,amssymb}

\newcommand{\epigraph}[3]{\par
\hfill\parbox{0.6\textwidth}{\footnotesize #1 \par \hfil #2
\textit{#3}}\par}%

\newcommand{\bp}{{\bf P}}

\newcommand{\om}{\omega}
\newcommand{\beq}{\begin{equation}}
\newcommand{\eeq}{\end{equation}}

\title{Probabilistic pathway representation of cognitive infromation}
\author{Andrei Khrennikov\footnote{Supported in part by the EU Human
Potential Programme, contact HPRN--CT--2002--00279 (Network on
Quantum Probability and Applications to Physica and Biology).} \\International Center for Mathematical
Modeling \\ in Physics and Cognitive Sciences,\\
University of V\"axj\"o, S-35195, Sweden\\
Email:Andrei.Khrennikov@msi.vxu.se
}
\date{}

\begin{document}
\maketitle

\epigraph{One held that psychological functions such as language or memory could never be traced
to a particular region of brain.
If one had to accept, reluctantly, that the brain did produce the mind, it did so as a whole and not as a collection
of parts with special functions. The other camp held that, on the contrary, the brain did have specialized parts and
those pars generate separate mind functions.}{Antonio R. Damasio}.

 \begin{abstract} We present for mental processes
the program of mathematical mapping which has been successfully realized
for physical processes. We emphasize that our project
is not about mathematical simulation of brain's functioning as a complex physical system,
i.e., mapping of physical and chemical processes in the brain on mathematical spaces.
The project is about mapping of purely mental processes on mathematical spaces.  We present
various arguments -- philosophic, mathematical, information, and neurophysiological --
in favor of the $p$-adic model of mental space. $p$-adic spaces have structures of hierarchic trees
and in our model such a tree hierarchy is considered as an image of neuronal hierarchy.
Hierarchic neural pathways  are considered as fundamental units of information processing.
As neural pathways can go through whole body, the mental space is produced
by the whole neural system.  Finally, we develop  Probabilistic Neural Pathway Model
in that Mental States are represented by probability distributions on mental space.
\end{abstract}

\section{Introduction}
In this paper (see also [1]-[10]) we present for mental processes
the program of mathematical mapping which has been successfully realized
for physical processes.
Of course, this is the project of huge complexity and it would be naive to
expect to realize it in one paper or even in a series
of papers. We emphasize that our project
is not about mathematical simulation of brain's functioning as a complex physical system,
i.e., mapping of physical and chemical processes in the brain on mathematical spaces.
The project is about
mapping of purely mental processes on mathematical spaces. In particular, the problem of the greatest
importance is the creation of an adequate mathematical model of {\it mental space.}
We agree that one can say that the project was started two thousands years ago by famous Greek
thinkers, Socrates, Plato, Aristotle. So we can say that our paper returns us to the old question
of {\it whether  reality, either material or mental, may be fundamentally mathematical.} The question about
material reality was answered in the positive. The question about mental reality is still open.
While harking back to Greek thought on the possibly mathematical nature of mental reality,
the project is unusual from a contemporary point of view. These days, the more usual role of
mathematics lies in investigation of biological, chemical, and physical bases of mental processes
(even including the possibility of their quantum origin). Our project shifts the focus to the mathematical
modeling of pure mental, rather than physical and chemical processes. It is hard to forsee whether
or not this strategy will be successful. Some form of our mathematical modeling may well prove
effective in psychology, psychiatric treatment and so forth. The crucial point is that we apply
{\it a new mathematics based on so called $p$-adic numbers} and it is too early to say whether
it is adequate to mental modeling. It is nevertheless possible that $p$-adics will open
new paths in mapping of mental processes on mathematical spaces. The main reason for thinking that $p$-adics may
be of value is that they appear better suited than mathematics based on real
or complex numbers to the heterogeneous complexity and discontinuity of mental processes.

If $p$ is a prime number (and this case is the most interesting from
the mathematical viewpoint) then $p$-adic numbers, like real, form what mathematicians
term a `field.' They can be added, subtracted, multiplied or divided without leaving the set of $p$-adics
(for any given $p).$ This is the case with the sets of rational and complex numbers too, but not with
all infinite sets of numbers. Integers, for example, do not form a `field' since dividing one by another can lead
to a fraction outside the set of integers. In such a case mathematicians use a
term `ring.' The set of $p$-adic numbers for non prime $p$ is not a field, but just a ring. Geometrically
$p$-adic rings are represented by hierarchic trees and this is one of the fundamental
features of $p$-adics used in mathematical modeling of mental processes. On $p$-adic trees
there are defined a $p$-adic metrics which are induced by the hierarchic structures of trees.
These are so called {\it ultra-metrics.} That $p$-adic spaces are metric spaces and yet are different from
`real' metric spaces is crucial for our project. Their metrics allow transfer of the mathematical
machinery developed for `real' spaces to weird-seeming spaces that are potentially better suited to
describing mentality.

We recall the main steps in mathematical mapping of physical processes and then
we will discuss the possibility to apply this scheme to mental processes.

\section{Mapping of physical processes on mathematical spaces}

{\bf Step 1. Elaboration of the mathematical model of physical space.}
At the first stage of development of physics (which at that time was not separated from philosophy)
there was created a mathematical model of physical space. This was the crucial step,
because before to start the study of dynamics of material systems we should
have some space which ``contains'' these systems.\footnote{Such a point of view was dominating in physics;
in philosophy it was strongly supported by Kant.} This stage of development took
a few thousands years. First of all it should be  mentioned the great contribution of ancient Greek mathematicians
and philosophers. I would like to pay attention to the works of Euclid and Aristotle.
Everybody knows that the Euclidean geometric model was the first axiomatic mathematical
model of physical space. But not so many people know about ideas of Aristotle
on the structure of physical and mental (!) spaces. We shall come back to Aristotle's
ideas on geometry of mental space in section 4 and now we discuss only models of physical space.
It seems that he was the first who presented a detailed discussion on {\it{continuity}}
of physical space. For him physical space (in which material objects are represented) is infinitely divisible
and continuous in the sense that it can not be represented as union of
two parts which do not have common boundary.

These ideas were developed by Newton and (especially) Leibnitz and, finally,
there was created (through works of 19th century mathematicians Cantor and Dedekind)
the modern model of physical space - {\it{real continuum.}} One dimensional
continuum is geometrically represented by the straight line. For further comparative
analysis it is important to underline that the straight line is an {\it{ordered}} set:
for any two points $x$ and $y$, we can say that $x \leq y$ or $y \leq x.$
We also recall that continuum has the algebraic structure of the {\it{field}} of
real numbers.

We also should mention the great invention of Descartes who introduced {\it cartesian systems of
coordinates} in physical space. We present the well known story about Descartes' discovery
of systems of coordinates. This story is not just curious, but it would play an important role
in the comparative analysis of mapping of  physical and mental processes on mathematical spaces.

{\small Once Descartes stayed in a hotel and he occasionally looked through
a window and saw a {\it{tree}} in the garden nearby.  There was a metal lattice on the window.
And that was the point! Descartes imagined the encoding of various parts of the tree by using the lattice. He understood
that by using lattices with smaller and smaller cells
one could create better and better encodings of the root, the trunk,  branches, and leaves of the tree.}

{\bf Step 2. Dynamical equations.}
Material objects  were mapped on real continuum (to be more precise: on the cartesian
product of  three real lines) and it became possible to describe motions of
such objects in this mathematical space.

{\bf{Step 2a. Second Newton law.}}
I. Newton formulated the fundamental dynamical law:
\[m\;a\;s\;s\;\;\;\times \;a\;c\;c\;e\;l\;e\;r\;a\;t\;i\;o\;n\;=\;f\;o\;r\;c\;e\]
Since the acceleration $a=\frac{d^2x}{dt^2} (t),$ where $x(t)=(x_1(t), x_2(t), x_3(t))$
is the trajectory of a physical system,
this law can be written as a differential equation:
 \beq
 \label{NE}
 m \frac{d^2x}{dt^2} (t)=f(x),
\eeq
where $m$ is the mass and $f(x)$ is the force acting to a system. By fixing the initial position and
the velocity, $x(0)=x_0, \; v(0)=\frac{dx}{dt}(0)=v_0,$
we determine the trajectory $x(t)$ as the unique solution of (\ref{NE}).
Therefore Newtonian mechanics is a deterministic theory: if initial conditions are fixed
the trajectory is uniquely determined.

{\bf{Step 2b. Hamiltonian equations.}} An important reformulation of Newton's mechanics is
given by {\it{Hamiltonian formalism.}}
Let us introduce the {\it momentum}  $\xi=m v \equiv m \frac{dx}{dt}$ and the
{\it energy}
$$
 H(\xi, x)=\frac{\xi^2}{2m} + V(x),
 $$ here $H_{\rm{Kin}}=\frac{\xi^2}{2m}$ and $H_{\rm{potential}}=V(x)$ are {\it kinetic and potential
energies.} A force $f$ is said to be {\it{potential}} if $f(x)=-\frac{dV}{dx}(x).$
For potential forces $f,$ the Newton equation (\ref{NE}) can be rewritten in the form
\beq
\label{H}
\frac{dx}{dt}=\frac{\partial H}{\partial \xi},\; \; \frac{d\xi}{dt}=-\frac{\partial H}{\partial x}
\eeq
The first equation is just the definition of the momentum $\xi$ and the second coincides with (\ref{NE}).
The system (\ref{H}) of Hamiltonian equations with initial conditions
$x(0)=x_0, \xi(0)=\xi_0$
determines uniquely the trajectory $(x(t), \xi(t))$ in the so called {\it{phase space}} - the
cartesian product of 3 + 3 real lines. This is the classical phase space dynamics.

{\bf{Step 3: Statistical mechanics.}}
Consider millions of particles which
motions are described by the Newton (or Hamiltonian) equations. Existence of such a
model  is extremely important from the philosophic
viewpoint, but this deterministic  model is not so useful for applications. It is meaningless
to describe mathematically
millions trajectories. Moreover, by investigating  individual trajectories we cannot find some
``collective characteristics'' of an ensemble of particles such as, e.g., energy, temperature.
In this situation it is natural  to describe statistical behavior of ensembles of particles.
Let us consider the simultaneous probability distribution of particles' position $x$ and momentum
$\xi,$ and denote the density of this probability distribution by $\rho(x, \xi).$ Thus the probability
to find a particle in a domain $O$ of the phase space can be calculated as
$$
 \bp((x, \xi) \in O )=\int \int_O \rho(x, \xi) dxd\xi,
 $$
 By using the Hamiltonian equations it
 is easy to derive the evolution equation for the density $\rho.$ This is the {\it Liouville equation:}
\beq
\label{LE}
\frac{\partial \rho}{\partial t} =\{\rho, H\} ,
\eeq
where $\{\rho, H\}$ is the Poisson bracket of functions $\rho$ and $H:$
$$\{\rho, H\}=\frac{\partial \rho}{\partial x} \frac{\partial H}{\partial \xi}-
\frac{\partial \rho}{\partial \xi} \frac{\partial H}{\partial x}.
$$
The probability distribution $\bp$ on the phase-space is the basic object of statistical physics.
One can forget about behavior of individual systems and investigate only probabilities.

{\bf{Step 4. Stochastic processes, Brownian motion, diffusion.}}
Let us consider the motion of a Brownian particle. The trajectory of such a particle can be extremely
irregular. Almost all trajectories are not smooth.
Thus the basic tool of Newtonian mechanics --
differentials -- becomes totally meaningless. It is impossible to apply the Newton
(or Hamiltonian) equation to describe the dynamic of a particle.
However, there were developed new mathematical methods based on {\it{stochastic differential equations}}
which give the possibility to describe dynamics
of Brownian-like systems:
\beq
\label{SE}
dx(t, \omega)=a(x(t, \omega)) dt + b(x(t, \omega)) d\omega(t),
\eeq
where $dt$ is the ordinary differential and $dw(t, \omega)$ is the Ito
differential -- an ``infinitesimal element of the Brownian process.''
Here the parameter $\omega$ describes  ``chance.''
It is natural to consider $\om$ as the label for a particle: $x(t,\om)$ is the trajectory
of the particle $\om$. This stochastic differential equation completed by the initial condition
$x(0, \om)=x_0(\om)$
describes the dynamics of a particle $\omega.$ At the first sight there is no difference with deterministic
Newtonian mechanics. But, in fact, the difference is very large. The crucial point is that a solution
of a stochastic differential equation is unique (again for Lipschitz coefficients)
only up to {\it stochastic equivalence.} We recall that two stochastic processes $x(t, \om)$ and $y(t, \om)$
are stochastically equivalent if for any $t:$
\beq
\label{SE2}
{\bf P} (\Omega_t)= 0, \; \mbox{where}\;
\Omega_t=\{\omega: x(t, \om)\not= y(t, \om)\}.
\eeq
Two solutions $x(t, \om)$ and $y(t, \om)$ of the same  stochastic differential equation
(with the same initial condition) can
be different, but only with probability zero (so for negligibly small number of particles).
However, these negligibly small sets of particles $\Omega_t$ depend on $t.$
Thus, for a time interval $[0,\delta],$
solutions can be essentially different. Therefore in theory of stochastic differential equations
people are interested not in individual solutions, but in corresponding
probability distribution $\bp(t, U)=\bp(\om:x(t,\om) \in U).$ This is the
probability that at the instant of time $t$ a particle $\om$ belongs to a domain $U$
of physical space. The density of this probability distribution ${\bf p}(t, x)$ satisfies
the Kolmogorov's forward equation\footnote{This equation is known to physicists as the Fokker-Planck
(or continuity equation).}:
\begin{equation}
\label{CK}
\frac{\partial {\bf p}}{\partial t}(t,x)=L {\bf p}(t,x), \lim_{t \downarrow 0}{\bf p}(t,x)={\bf p}_0(x) ,
\end{equation}
Here $L$ is the generator of the evolution of probability distributions. We consider a particular class
of generators $L$ which in the one dimensional case have the form:
$$
L{\bf p}(x)= \frac{1}{2}\frac{d^2 }{d x^2}[b^2(x) {\bf p}(x)] - \frac{d}{d x} [a(x) {\bf p}(x)].
$$
The corresponding Kolmogorov's forward equation (\ref{CK}) describes diffusion and the corresponding
stochastic differential equation is called {\it{diffusion equation.}}\footnote{The coefficients
$a(x)$ and $b^2(x)$ are called drift and diffusion coefficients. By neglecting inertia of a particle
it can be assumed that its motion consists of two components: drift induced by macroscopic velocity of flow
of liquid and fluctuations induced by chaotic motion of molecules of liquid.}
In particular, if $b=1$ and $a=0,$ i.e. $L = \frac{1}{2}\frac{d^2}{d x^2},$
the equation  (\ref{CK}) describes the evolution of the probability distribution
for Brownian motion (Wiener process).

{\bf{Step 5: Quantum physics}}
We shall not consider quantum models in this paper, see, e.g., [10]
and the extended bibliography in that paper. Quantum mechanics is also a statistical theory:
all quantum experiments are about statistical behavior of huge ensembles of quantum particles.
However, in the opposition of classical statistical physics, it is assumed (at least by those
who use the orthodox Copenhagen interpretation) that quantum probabilities cannot be reduced
to ordinary ensemble probabilities. Thus there is no any deterministic prequantum process
-- neither deterministic (as in classical statistical physics) nor stochastic (such as the diffusion
process)  -- which could generate quantum probabilistic distributions.\footnote{Of course, there
are some attempts to construct such processes, e.g., Bohmian mechanics (deterministic prequantum process),
or stochastic electrodynamics (prequantum stochastic process).}

\section{Mapping of mental processes on mathematical spaces}
We repeat for mental processes the scheme of
mapping on mathematical spaces which has been successfully realized in physics:

{\bf{Step 1: Elaboration of the notion of mental space.}}
We should find an adequate mathematical model of {\it{mental space}} $M.$
Points of $M$ (mental points) will represent {\it elementary mental states.}

{\bf{Step 2: Dynamical equations in mental space.}} There should be presented
dynamical equations describing trajectories $x(t)$ of elementary states in the mental space $M.$

{\bf{Step 3: Mental statistical dynamics.}} In the same way as in classical statistical physics it
is natural to study not behavior of individual states
(millions of trajectories in $M),$ but probabilistic behavior
of huge statistical ensembles of elementary states. And this is not just the question of simplification
of the mathematical description. There are numerous evidences that the mind is the product of the
activity of the whole brain, see, e.g., [11]-[13].
 Thus mind should be produced by ensembles and not individual states.
Later we shall present some neurophysiological arguments in favor of ``ensemble-mind.''
At the moment we  present a general argument from information theory. This is the argument
of {\it security of processing of information.} For an information system producing
millions of  states it would be very dangerous to realize functions of large
importance by operating with individual states. There should be developed the ability
to take ``collective decisions.''
Probability is the fundamental factor determining such decisions.
On the other hand, we can not exclude that some simple mental functions can be realized
through dynamics of individual states.

{\bf{Step 4. Mental diffusion.}}
As we know from physics, classical statistical mechanics is still deterministic. There is the perfect
mechanical order described by the Newton (or Hamilton) equations. Trajectories are {\it{smooth}} and
particles do not perform actions violating Newtonian order! Intuitively the mental dynamics does not look
so well ordered as the Newtonian dynamics. It may be more natural to consider in mental space
$M$ the stochastic dynamics,  instead of the deterministic dynamics. Later we shall present a neurophysiological argument
supporting the stochastic dynamical model (in mental space!).

We emphasize that by choosing the stochastic
mental dynamics, instead of the Newtonian mental dynamics, we do the step having fundamental consequences
for cognitive science. Assume that the mental behavior is determined only by probabilities.
Then different (stochastically equivalent) processes $x(t, \om), y(t, \om), \ldots$ (which satisfy the same
stochastic differential equation in the mental space $M$ with the same initial condition)
induce the same mental  behavior (determined by the probabilities).

In our model (see section 6) elementary mental states are produced by neural structures. In this
model stochastic evolution of elementary mental states implies the violation of the
materialistic axiom of cognitive science. In our model:

\medskip

{\it{Different  stochastic neural dynamics can produce the same mental behavior.}}

\medskip

It may be that the materialistic axiom is violated only for high level mental functions which operate with
probabilistic distributions produced by neuronal ensembles.

This is our program of a mathematical mapping of mental processes on mathematical spaces.
It seems that precisely in this form the program has never been formulated. Nevertheless, there were
developed various approaches which  realized some steps of this program (of course,
the same step can have  a few different realizations depending on choices of mathematical models for mental
space and dynamical equations).
We mention {\it{Dynamical System Approach}} [14]-[19] which is based on the following principles:

\medskip

(a) {\it Embodiment} of mind;

(b) {\it Situatedness} of cognition.

We recall that the orthodox Dynamical System Approach emphasizes commonalties between
behavior in neural and mental processes on one hand and with physiological and environmental events on the other.
The most important commonality is the dimension of time shared by all these domains. In fact, sharing of
time scales  with physical environment implies

(c)  Continuous real time evolution described by differential equations.

\medskip

At the first step people working in the DS-approach decided to borrow the model of space from
physics.
This was an attempt to embed mind into the Euclidean physical space. This step (the special choice of space)
determined the following development of the DS-approach. At the second step there were applied
standard differential
equations describing physical processes in the brain as a physical system. In particular, there are widely
applied oscillator models.

We can also mention {\it symbolic models} [20], [21], and {\it{Neural Network Approach}}, [22]--[25]. In the
opposition of the DS-approach, in the NN-approach
physical space does not play an important role. The fundamental role is played by {\it potentials}
$x_i(t)$ corresponding to neurons  $i=1,...,N.$ But this using of physical potentials also implies
application of the standard mathematical models of physics. ``Mental space'' of the NN-approach --
the space of potentials -- is also the real space which is the  mathematical basis of physics.\footnote{We remark
that in the NN-approach there were realized main steps of our program, including stochastic neural models.}

\section{p-adic hierarchic mental space}
In [1] there was proposed the $p$-adic model of mental space. We shall give the detailed
description of the $p$-adic mental space later and now we discuss some general arguments
in favor of this choice.

{\bf a.) Hierarchic structure of mental space.} In neurophysiology and psychology
there is intensively discussed the idea that mental processes have hierarchic structures.
I would like to say that the mental image of the physical world is a special hierarchic representation
of this world.
Thus a mental space should be endowed with a hierarchic structure which would adequately represent
hierarchies observed in psychology and cognitive science.
However, the real continuum is totally homogeneous; there is no natural hierarchic structure on the real
continuous space. All physical points ``have equal rights'' and translation as well as rotation
invariance  are the fundamental features of all physical theories.
This is one of reasons to reject the real continuous model as a candidate to a model of mental space.
On the other hand, the $p$-adic space has a natural hierarchy which is induced by {\it{a tree structure}}
of this space.

{\bf{b.) Tree structure, mental space as a product of neuronal activity.}}
It is clear that mental space should be coupled  with the material structure of the brain.
This structure should be represented in some way in mental space.\footnote{This statement is not about
{\it reductionism.} We do not claim that mind can be directly reduced to neuronal activity. As was mentioned,
in our model the materialistic axiom of cognitive science is violated. We just pay attention to the fact that
physical neural and mental structures should be coupled.  We even need not stay
on the materialistic position. One could not totally exclude the possibility that material neuronal
structures are images of mental structures which are located in mental space. However, I would not like to
go deeply in such philosophic problems. This paper is simply an attempt to create a mathematical model
which contains new mental coordinates which differ essentially from physical real coordinates.}
It is well known that many neuronal configurations in the
brain have the tree-structure. It seems to be natural to have such a tree structure on mental space.
And we remark that $p$-adic spaces have tree structures,
see Figure 1 - the tree representation of the 2-adic space.
\begin{center}
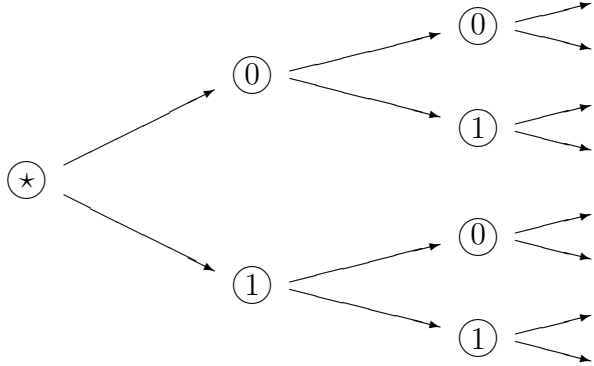
\begin{figure}[ht]
\unitlength1cm
\begin{picture}(12,5)
\put(2,3){\circle{0.5}}            \put(1.9,2.9){$\star$}
\put(2.5,2.8){\vector(2,-1){2.0}}
\put(2.5,3.2){\vector(2,1){2.0}}
\put(5,4.4){\circle{0.5}}            \put(4.9,4.28){$0$}
\put(5.5,4.35){\vector(4,-1){2.0}}
\put(5.5,4.45){\vector(4,1){2.0}}
\put(5,1.6){\circle{0.5}}            \put(4.9,1.48){$1$}
\put(5.5,1.55){\vector(4,-1){2.0}}
\put(5.5,1.65){\vector(4,1){2.0}}
\put(8,5.05){\circle{0.5}}          \put(7.9,4.95){$0$}
\put(8,3.7){\circle{0.5}}          \put(7.9,3.57){$1$}
\put(8,2.25){\circle{0.5}}          \put(7.9,2.15){$0$}
\put(8,0.9){\circle{0.5}}          \put(7.9,0.75){$1$}
\put(8.5,5.00){\vector(4,-1){1.0}}
\put(8.5,5.10){\vector(4,1){1.0}}
\put(8.5,3.65){\vector(4,-1){1.0}}
\put(8.5,3.75){\vector(4,1){1.0}}
\put(8.5,2.20){\vector(4,-1){1.0}}
\put(8.5,2.30){\vector(4,1){1.0}}
\put(8.5,0.85){\vector(4,-1){1.0}}
\put(8.5,0.95){\vector(4,1){1.0}}
\end{picture}
\caption{The $2$-adic tree}
\end{figure}
\end{center}
For an arbitrary natural number $p > 1,$ the $p$-adic tree is constructed in the same way:
there are $p$-branches leaving each vertex. In principle, we can consider trees,
where the number of branches depends on a vertex.
However, mathematics on such trees is essentially more complicated, cf. [1].

We now turn back to the story about the discovery of the cartesian system of coordinates.
{\small  Suppose for a moment that Descartes paid attention not to the lattice on the window of his room,
but to the internal hierarchic structure of the tree: the root, the trunk, branches, subbranches, leaves.
In such a case he would discover not the ordinary orthogonal system of coordinates,
but a $p$-adic hierarchic system on the tree.}

{\bf{c.) Absence of linear order in the space of mental states.}}
As we know, the real continuum is a {\it linearly ordered set.}
Any set of points $x_1, \ldots, x_m$ on the straight line can be ordered.
On the other hand, there are no reasons to assume that such an order structure can
be introduced on the space of mental states. It is impossible to
order all minds, feelings, emotions, ....; for example, it is very doubtful that
it would be possible to create  a model of mental space such that for any two minds
$x$ and $y$ it would be possible to say $x \leq y$ or $y \leq x.$
Therefore the absence of a linear order on a $p$-adic tree
is an attractive feature of the $p$-adic model which can be useful in
mathematical modeling of mental processes.

{\bf (d.) Link to Aristotle, ultrametricity.} As was already remarked, Aristotle discussed in detail
geometries of physical and mental spaces (for him the latter was associated  with the space of words),
see [26]. Continuity was considered as the main distinguishing feature of physical space. For Aristotle
the term continuity was related not to functions on some spaces (as mathematicians do in the
contemporary analysis),
but to spaces by themselves. Aristotle's
notion of {\it continuous space} coincides with the notion of {\it connected space} which is used in
contemporary mathematics. A space is said to be connected if it is impossible to split it
into two parts in  such a way that
there are no boundary points. Aristotle presented important arguments that physical space is
continuous-connected.
At the same time he underlined that mental space (space of words) is discontinuous (disconnected).
 Such a space
can be split into two parts without boundary points.

By using the contemporary terminology one can say that two thousands years ago there was
discussed the idea that mental spaces should be represented as
disconnected mathematical spaces. The class of disconnected topological spaces is well investigated,
[27], [28]. We pay attention to the important subclass
of disconnected spaces -- spaces which are endowed with metrics. This is a purely mathematical restriction
which essentially simplifies modeling (but in principle there may exist
natural mental applications of non metrizable disconnected mental spaces). Under some mathematical
restrictions we get the class of so called
{\it ultrametric spaces.} We recall that a metric is called ultrametric if
it satisfies the {\it strong triangle inequality}:
$$
\rho(x,y) \leq \max[\rho(x,z), \rho(z,y)] .
$$
The strong triangle inequality can be stated
geometrically: {\it each side of a triangle is at most as long as the
longest one of the two other sides.}
Finally, we mention a theorem of general topology [28] which says that
{\it each ultramtric space can be represented
as a tree and vice versa.} So we come to the tree model of mental space which was developed in [1]-[10]. Such a model
can be considered as a modern version of the Aristotle's model of mental space.
I interpret ultrametricity
as the topological encoding of the hierarchic structure of a tree. By operating in an ultrametric space we
can forget about the underlying hierarchic structure and use the language of analysis:
neighborhoods, open and closed sets,
convergence of sequences, continuity of functions (the contemporary analysis gives the possibility to consider
continuous functions even on discontinuous spaces).\footnote{In many particular cases such a relation between ultrametricity and hierarchy was
used in theory of spin glasses, see, e.g., [29]-[31].}

{\bf (e.) $p$-adic trees and numbers.} The class of general ultrametric spaces is still too large for mathematical
modeling (at least at the first stage of realization of our program).
In mathematical modeling of physics the important role is played not only by continuity (topology) of the space, but also
by the presence of an algebraic structure on this space; elements of real line can be interpreted as numbers
(real numbers). We would like to have such a structure on mental space.
We need ``mental numbers.'' There is well defined algebraic structure
on a special class of trees, so called $p$-adic trees. These are trees where $p$-branches leave each vertex
and $p$ is a prime number. This number is an invariant of a tree and it does not depend on a vertex.
As was already  remarked, on such a tree we can introduce the structure of a field;  if $p$ is not a prime
number, then the  $p$-adic tree is only a ring, [1], [27], [32], [33].
By using Figure 1 we can encode infinitely long branches of the 2-dic tree by strings
of zeros and ones. Thus instead of branches of  a tree we can work with such strings.
In the general case of a $p$-adic tree
these are strings of symbols belonging to the set $\{0,1,..., p-1\}.$
The distance between two strings $x$ and $y$ representing branches of  the $p$-adic tree is defined as
$$
\rho_p(x,y) =\frac{1}{p^{l(x,y)}},
$$
where $l(x,y)$ is the length of the common left-hand side part of strings $x$ and $y.$
This is an ultrametric. So any $p$-adic tree is an ultrametic space. For example, let $p=2$ and
let us consider
strings $x=(000...)$ and $y=(001...).$ Then $l(x,y)=2$ and $\rho_2(x,y)= 1/4.$
Let $z=(01...).$ Then $l(x,z)=l(y,z)=1$ and hence $\rho_2(x,z)= \rho_2(y,z)=1/2 > 1/4=\rho_2(x,y).$

The set of $p$-adic numbers is denoted by the symbol ${\bf Q}_p.$
We choose ${\bf Q}_p$ as a mathematical model of mental space.
In more general models we consider mental spaces
which have many $p$-adic coordinates: cartesian products of a few $p$-adic trees;
or even trees corresponding
to different numbers $p_1,p_2,...,p_n.$ In our model $p$-adic numbers are considered as {\it mental numbers.}
At the moment we do not assign some special cognitive meaning
to branches of $p$-adic trees. There can proposed various models. In a series of papers [2], [5], [6] we
encoded {\it psychological states} by branches of $p$-adic trees. In this paper we would like
to find links to neurophysiology and associate mental strings with activity of neurons, see section 6.

{\bf (f.) $p$-adic dynamical models for mental processes.} We considered discrete dynamical systems
$$
x_{n+1}= f(x_n), n=0,1,...,
$$
corresponding to maps $f: {\bf Q}_p \to {\bf Q}_p, x \to f(x).$
We use the standard terminology of the theory of dynamical
systems; see, for example, [1].
{\small If $f(a)=a$ then $a$ is a {\it fixed point}.
If $x_n=x_0$  for some $n=1,2,...$ we say that $x_0$ is a {\it periodic
point}. If $n$ is the smallest natural number with this property then $n$
is said to be the {\it period} of $x_0.$
A fixed point $a$ is called an {\it attractor}
if there exists a neighborhood (ball) $V(a)$ of $a$ such that
all points $x_0 \in V(a)$ are attracted by $a,$ i.e.,
$\lim_{n \to \infty} x_n= a.$ }

In [1]-[8] there was proposed a dynamical model for the process of thinking
which is based on {\it hierarchic coding} of mental information by $p$-adic numbers and
processing by maps in $p$-adic trees. These are continuous maps with respect to tree's ultrametric.
We recall that the tree-hierarchy is encoded into the ultrametric topology. Thus such maps preserve
hierarchic encoding of information.

Another basic feature of our dynamical model [2] is that
the process of thinking is  split into two separate (but at the same time closely connected)
domains: the {\it conscious} and the {\it unconscious.} We used the following point of view
of the simultaneous work of the consciousness
and unconsciousness. The consciousness contains a control center {\it CC} which has
functions of control. It
formulates problems and
sends them to the unconscious domain. The process of finding a solution is hidden
in the unconscious domain. In the unconscious domain there works a gigantic
dynamical system. Its work starts with a mental state $x_0$ (or a group of mental states $U_0$)
which has been communicated by the consciousness. Mathematically it corresponds
to the choice of an initial point $x_0$ (or a neighborhood $U_0$). Starting with
this initial point $x_0$ a {\it thinking processor} $f$ in the unconscious domain
generates at tremendous speed a huge number of new
mental states:
$$
x_1=f(x_0),...., x_{n+1}= f(x_n),...
$$
These mental states are not used by the consciousness.
The consciousness (namely ${\it CC}$)
controls only some exceptional moments in the work of
the dynamical system in the unconscious domain. These are different regimes of
stabilization. First, there
are {\it attractors} which are considered by the consciousness as possible
solutions of the problem. Then there are cycles
$(a\to b\to \cdots\to c\to a)$ which generate
signals to stop the work of the dynamical system.
If the consciousness cannot take a decision
then it can
send  a new initial mental state $x_0^\prime$ to the unconscious domain  or change the regime of
work of a thinking processor in the unconscious domain. Mathematically
the change of a regime can be described as the change of a function
$f(x)$ which determines the dynamical system. Thus we can describe
the process of thinking as the work of a family  of dynamical systems
$f_\alpha(x),$ where the parameter $\alpha$ is controlled by
the consciousness (or chance in random dynamical thinking models, [6]).
This model was used for modeling of psychological behavior. In particular, in [2] we simulated
sexual behavior by encoding mental states of sexual partners by hierarchic strings of information:

{\small x=(sex, age, education level, ...)}

The sexual
dynamical system $f_{\rm{sex}}$ operates with such hierarchical images and produces ``solutions''
(or exhibit cyclic behavior).

The main problem of this model was impossibility to choose functions $f$ generating dynamics
by using purely psychological reasons. We were able only provide animation corresponding to simple functions
on $p$-adic trees and give psychological interpretation to results of such animation.
Another reason for looking for new models was that the above $p$-adic dynamical model was purely  deterministic. As was already mentioned,
it seems to be more natural to study models in which mental information is processed probabilistically; models
in which mental states are represented by fields of probability on mental space and not by single points of
this space. Finally, in [1]-[6] we were not able to couple hierarchic $p$-adic encoding of information with neurophysiology.
In following sections we try to solve this problem of mind-body relation and then  proceed to probabilistic
processing of mental information.

\section{Coupling of the $p$-adic mental space with the neural structure}
In previous sections there were presented some general arguments in favor of the
$p$-adic model of mental space. Now we present the neurophysiological  basis of the
$p$-adic mental space.  Coupling between the neuronal material structures and the $p$-adic
mental structures is based on the correspondence between the neuronal hierarchy (i.e.,
hierarchic relations between neurons and groups of neurons) and the $p$-adic tree hierarchy.

In our model each psychological function is based
on a graph of neural pathways, {\it cognitive graph,} that is {\it centered} with respect
to one fixed neuron. Thus basic units of processing of mental information  are {\it centered neural pathways}
and basic units of mental information are {\it centered  strings of firings} produced by centered
neural pathways.

\begin{figure}[ht]
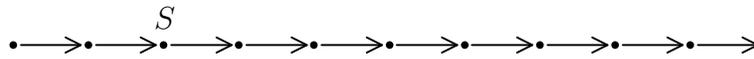

\begin{center}
  \begin{texdraw}
    \drawdim cm
    \move(0 0)\fcir f:0 r:.05
    \move(1 0)\fcir f:0 r:.05
    \move(2 0)\fcir f:0 r:.05
    \move(3 0)\fcir f:0 r:.05
    \move(4 0)\fcir f:0 r:.05
    \move(5 0)\fcir f:0 r:.05
    \move(6 0)\fcir f:0 r:.05
    \move(7 0)\fcir f:0 r:.05
      \move(8 0)\fcir f:0 r:.05
        \move(9 0)\fcir f:0 r:.05

                \arrowheadtype t:V
    \arrowheadsize l:.2 w:.2
    \move(0.1 0)\avec(0.9 0)
        \move(1.1 0)\avec(1.9 0)
         \move(2.1 0)\avec(2.9 0)
          \move(3.1 0)\avec(3.9 0)
           \move(4.1 0)\avec(4.9 0)
            \move(5.1 0)\avec(5.9 0)
             \move(6.1 0)\avec(6.9 0)
              \move(7.1 0)\avec(7.9 0)
               \move(8.1 0)\avec(8.9 0)
                \move(9.1 0)\avec(9.9 0)

    \move(2 0.2)\textref h:C v:B \htext{$S$}
  \end{texdraw}
\end{center}
\caption{Centered pathway}
\end{figure}

\medskip

Centering determines a {\it hierarchic structure} on the cognitive graph
and on the corresponding space of mental points.

\begin{figure}[ht]
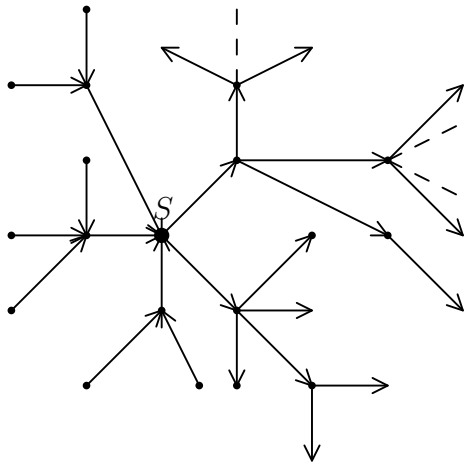

\begin{center}
  \begin{texdraw}
    \drawdim cm
    \move(0 2)\fcir f:0 r:.05
    \move(1 4)\fcir f:0 r:.05
    \move(1 5)\fcir f:0 r:.05
    \move(1 2)\fcir f:0 r:.05
    \move(1 0)\fcir f:0 r:.05
    \move(2.5 0)\fcir f:0 r:.05
    \move(2 1)\fcir f:0 r:.05
    \move(2 2)\fcir f:0 r:.1
      \move(3 0)\fcir f:0 r:.05
        \move(3 1)\fcir f:0 r:.05
          \move(3 3)\fcir f:0 r:.05
            \move(3 4)\fcir f:0 r:.05
              \move(4 0)\fcir f:0 r:.05

                \move(4 2)\fcir f:0 r:.05
                  \move(5 2)\fcir f:0 r:.05
                    \move(5 3)\fcir f:0 r:.05
                         \move(0 1)\fcir f:0 r:.05
                         \move(1 3)\fcir f:0 r:.05
                           \move(0 4)\fcir f:0 r:.05
    \arrowheadtype t:V
    \arrowheadsize l:.2 w:.2
    %vänsterkanten
    \move(0 2)\avec(1 2)
        \move(1 0)\avec(2 1)
        \move(2 1)\avec(2 2)
        \move(2.5 0)\avec(2 1)
         \move(1 3)\avec(1 2)
         \move(1 2)\avec(2 2)
           \move(0 1)\avec(1 2)
            \move(0 4)\avec(1 4)
            \move(1 4)\avec(2 2)
                \move(1 5)\avec(1 4)
                    \move(4 0)\avec(4 -1)
                    \move(4 0)\avec(5 0)
                    \move(3 1)\avec(3 0)
                    \move(3 1)\avec(4 0)
                    \move(3 1)\avec(4 1)
                    \move(3 1)\avec(4 2)
                    \move(2 2)\avec(3 1)
                    \move(2 2)\avec(3 3)
                    \move(3 3)\avec(5 2)
                    \move(3 3)\avec(5 3)
                    \move(3 3)\avec(3 4)
                    \move(3 4)\avec(2 4.5)
                    \move(3 4)\avec(4 4.5)
                    \move(5 3)\avec(6 2)
                    \move(5 3)\avec(6 4)
                    \move(5 2)\avec(6 1)

    \lpatt(.2 .2)
    \move(3 4)\lvec(3 5)
    \move(5 3)\lvec(6 2.5)
    \move(5 3)\lvec(6 3.5)
    \move(2 2.2)\textref h:C v:B \htext{$S$}
  \end{texdraw}
\end{center}
\caption{Cognitive graph}
\end{figure}

A centering neuron $S$ should not
be considered as a kind of grandmother neuron. It simply determines a system of
mental coordinates (corresponding to
the concrete psychological function) on the neural system of a cognitive system.
Of course, such a model with one neuron centering of a psychological
function is oversimplified.
Complex psychological functions should be based on a few cognitive graphs centered with respect to
an ensemble of neurons. The centering hierarchic structure on a cognitive graph can be mapped on the $p$-adic hierarchic structure
with the aid of the frequency coding having $p$ as the base of coding, see section 6.3, 6.4.
We call this approach {\it Neural Pathway Approach.} The essence of this approach is the view to
the $p$-adic mental space as the product of hierarchically ordered neural activity and frequency
$p$-coding.

\section{Model: thinking on a cognitive graph}

{\bf 6.1. Localization of psychological functions.}
One of the strong sides of Neural Pathway Approach is a new viewpoint to the problem of
localization of psychological functions. Since an elementary
unit of mental processing is represented by a centered neural pathway and
a pathway can go through various domains of brain and body, there is no localization of mental
functions in the Euclidean geometry of the physical space that is typically used to describe physical
brain and body. On the other hand, a psychological function can be localized in the  space of all pathways.
\footnote{In fact, this is a kind of hierarchic localization, - compare to A. Damasio: ``What determines the
contribution of a given brain unit to the operation of the system to which it belongs is not
just the structure of the unit but also its place in the
system. ... The mind results from the operation of each of the separate
components, and from the concerted operation of the multiple systems constituted by those separate
components,''
p. 15, [34]. In our model there is even no place for ``separate
components"; everything is unified from the beginning due to the pathway
representation of cognitive information.}

We have to distinguish the space $\Pi$ of all centered neural pathways (hierarchic chains of neurons)
in the physical brain and body
and the space $M$ of all possible mental points that can be produced by elements of $\Pi.$
In principle, a few distinct elements of $\Pi,$ centered neural pathways, can produce (at same instant of
time) the same point $x\in M.$

{\bf 6.2. Body$\to$mind field.} Firings of neurons throughout centered neural pathways of
the cognitive graph produce elementary
mental states (points) involved in the realization of
a psychological function.  We denote a psychological function by the symbol $f$  and
the cognitive graph of the $f$ by the symbol $\Pi_f.$
How can we describe mathematically the functioning of $f$?
There are various levels of the mathematical description.

At the basic level we should provide the
 description of `body$\to$mind' correspondence. This correspondence is described by a function
 \[\varphi:\Pi_f \to M,\]
that maps centered neural pathways into mental points produced by these pathways:
$z \in \Pi_f \to x=\varphi(z) \in M.$ We call the map $\varphi(z)$ {\it{body$\to$mind field.}}\footnote{Of course,
$\phi$ depends on the psychological function $f: \; \phi= \phi_f.$}
The psychological function $f$ generates the evolution of the field  $\varphi.$ Starting with the initial
field $\varphi_0(z), f$ produces time-dependent field $\varphi(t, z).$
We have to consider very important problem of interpreting of the evolution parameter,
`time', t; in particular, relation between physical and psychological time.
We shall discuss this problem in section 9. At the moment we restrict ourself to consideration of
the discrete time evolution: $t=t_n=0,1,2,\ldots$
 By taking into account the process of wholeness of thinking  we describe the functioning of $f$ by
 an integral operator with the kernel $K(z,y):$
 \begin{equation}
 \label{I2}
 \varphi(t_{n+1}, z)=\int_{\Pi_f} K(z, y) \varphi(t_n, y)dy,
\end{equation}
where integration is performed over the cognitive graph, $\Pi_f.$
We notice that neither the space of
pathways $\Pi_f$ nor the space of mental points $M$ have
the Euclidean geometry. In particular, we cannot use
the ordinary real analysis to describe this model mathematically.
We should use the so called ultrametric analysis [1],[27], [32].\footnote{
The form of the kernel $K(z,y)$ is determined by the psychological function $f.$ We notice that
mental evolution (\ref{I2}) is represented by a linear integral operator in the space of body$\to$mind fields.
In principle, we can consider more general, nonlinear models. However, the model with summation over
the whole graph with a weight function $K(z, y)$ looks very natural.}
Thus we propose the following model of thinking:

\medskip

Each psychological function $f$
is based on a  graph of neural pathways, cognitive graph $\Pi_f,$
see, e.g., Figure 2. The cognitive graph has the hierarchic structure corresponding
to  the central neuron, $S,$ of this graph. The elementary unit of
mental information -- elementary mental state (or mental point) -- is given by the string of firings of neurons throughout a pathway, a branch of the graph.
There can be proposed various neural pathway coding systems based on strings of firings.

\medskip

{\bf 6.3. Firing/off (2-adic) coding.}  For each instant of time $t$, we assign to a neuron in
a pathway 1, if the neuron is firing, and 0, otherwise. Mathematically a mental point
is represented by a sequence of zeros and ones. Each sequence is centered with respect to the
position corresponding to firings of the central neuron $S.$

\medskip

Let us consider the geometric structure of the mental space $M$ corresponding to firing/off coding.
Here each centered pathway produces  a centered sequence of
zeros and ones. The most important digit, $x_0=0$ or 1, in a sequence $x \in M$ gives the state
of the central neuron $S.$
Hierarchy on the mental space is based on the exceptional role that is played by the central neuron.
This hierarchy induces the 2-adic topology on the mental space.

In mathematical modeling it is convenient to consider infinitely long neural pathways
and corresponding information strings, mental points (this is of course just a mathematical
idealization). In section 4 we introduced the 2-adic  metric for mental points produced by
cognitive graphs having only output  (with respect to the cental neuron $S)$ neural pathways.
Now we introduce the 2-adic metric for mental points produced by
general cognitive graphs which contain input as well
as output neural pathways.  The 2-adic distance $\rho_2$ is defined in the following way.
Consider two mental points:
$$
x=(... x_{-l}...x_0...x_k...)\; \mbox{and}\;
y=(... y_{-l}...y_0...y_k...), \; x_{\pm j}, y_{\pm j} = 0, 1 .
$$
We use index 0 for the state $x_0$ of the central neuron $S,$  negative indexes for states of  neurons that
produce inputs propagating to the $S$ through the ordered neural chain,
positive indexes -- for states of  neurons that are receivers of $S$-output.
First suppose that all input states coincide: all $x_{-l}= y_{-l}.$ Let $l \geq 0$ be the first
index such that $x_l \not= y_l.$ Then by definition
\begin{equation}
\label{PM}
\rho_2(x,y) = \frac{1}{2^l} .
\end{equation}
This is the distance considered in section 4. Thus if two neural pathways $z$ and $w$ produce strings $x$ and $y$ having the same
input part, then $\rho_2(x,y)$ goes to zero if the length $l$ of the common output part
goes to infinity.

Suppose now that input parts are different. Let $l$ be the first index (if we go from the left hand
side)  such that $x_{-l}\not= y_{-l}.$ Then by definition
\begin{equation}
\label{PM1}
\rho_2(x,y) = 2^l .
\end{equation}
Here larger common initial input part implies shorter distance between mental points.
Consider  two neural pathways starting at e.g. a sensory receptor.  Suppose that states
(e.g. on/off) of initial neurons in these pathways are distinct.  Then the distance
between corresponding mental points is large. It is important to remark that in such a situation
longer pathways induce larger distance between mental points, see also section 9.

We remark that a cognitive graph $\Pi_f$ producing the 2-adic mental space need not have a tree-structure.
Such a graph can contain numerous cycles. However, the corresponding mental space $M$ created by
the 2-adic coding always has the structure of the 2-adic tree.

{\bf 6.4. Frequency ($p$-adic) coding.} General $p$-adic coding (where $p\geq 2$ is a natural number)
may be induced by the
frequency coding.
We assign to each neuron in a pathway the frequency of firings. Frequencies of firing
are a better basis for the description of processing of information by neurons than
a simple on/off. This has been shown to be the fundamental element of neuronal communication
in a huge number of experimental neurophysiological studies (see e.g. [35], [36] on mathematical modeling
of brain functioning in the frequency domain approach). In the mathematical model it
is convenient to consider a discrete set of frequencies: $0, 1, .., p-1,$ where $p$ is some natural number.
Here frequency is the number of output spikes produced by a neuron during some unit of
psychological time (some period $\Delta$ of physical time, see section 6 for detailed consideration).
Thus mathematically a mental point is represented by a sequence of numbers belonging to the set
$\{0, 1,  \ldots, p-1\}.$ Information is not homogeneously distributed throughout such sequences.
The presence of the central neuron $S$ in the cognitive graph $\Pi_f$
induces a hierarchic structure for elements of an information sequence.

This system of neural pathway frequency coding
should be justified  by neurophysiological studies.
However,  at the
moment there are no experimental technologies that would give the possibility
to measure firings of neurons  throughout even one long pathway of individual neurons.
To confirm our pathway-coding hypothesis, we have to measure simultaneously firings of neurons
for a huge ensemble of neural pathways.

We underline that the system of coding and not the topological  structure of a cognitive graph determines
the structure of the corresponding mental space, see also section 9. Totally different cognitive graphs can produce the same
mental tree. For example, let us consider 2-adic coding. The graphs a,b,c on Figure 4 produce the same,
2-adic, mental space.

\begin{figure}[ht]
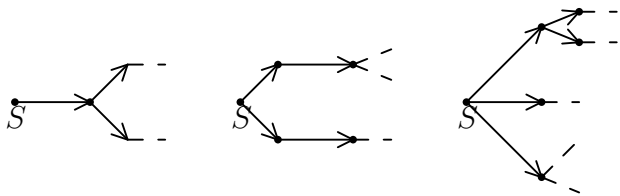

\begin{center}
  \begin{texdraw}
    \drawdim cm
  %4a
  %punkterna
  \move(0 0)\fcir f:0 r:.05
    \move(1 0)\fcir f:0 r:.05

    \arrowheadtype t:V
    \arrowheadsize l:.2 w:.2
    %pilarna
    \move(0 0)\avec(1 0)
     \move(1 0)\avec(1.5 0.5)
      \move(1 0)\avec(1.5 -0.5)

       %4b linjerna
       \arrowheadtype t:V
         \arrowheadsize l:.2 w:.2
         %pilarna
         \move(3 0)\avec(3.5 0.5)

         \move(3 0)\avec(3.5 -0.5)
          \move(3.5 0.5)\avec(4.5 0.5)
           \move(3.5 -0.5)\avec(4.5 -0.5)

     %4c linjer
     \arrowheadtype t:V
         \arrowheadsize l:.2 w:.2
         %pilarna
         \move(6 0)\avec(7 0)
          \move(6 0)\avec(7 -1)
           \move(6 0)\avec(7 1)
           \move(7 1)\avec(7.5 1.2)
           \move(7 1)\avec(7.5 0.8)

    \lpatt(.2 .2)
    \move(1.5 0.5)\lvec(2 0.5)
    \move(1.5 -0.5)\lvec(2 -0.5)

          \move(0 -0.35)\textref h:C v:B \htext{$S$}

         %4b från 3

          %punkterna
       \move(3 0)\fcir f:0 r:.05
         \move(3.5 0.5)\fcir f:0 r:.05
          \move(3.5 -0.5)\fcir f:0 r:.05
           \move(4.5 0.5)\fcir f:0 r:.05
            \move(4.5 -0.5)\fcir f:0 r:.05

         \lpatt(.2 .2)
         \move(4.5 0.5)\lvec(5 0.7)
         \move(4.5 0.5)\lvec(5 0.3)
         \move(4.5 -0.5)\lvec(5 -0.5)

               \move(3 -0.35)\textref h:C v:B \htext{$S$}

         %4c från 6

           %punkterna
  \move(6 0)\fcir f:0 r:.05
    \move(7 1)\fcir f:0 r:.05
          \move(7 -1)\fcir f:0 r:.05
              \move(7.5 0.8)\fcir f:0 r:.05
                  \move(7.5 1.2)\fcir f:0 r:.05
                      \move(7 0)\fcir f:0 r:.05

    \lpatt(.2 .2)
    \move(7 -1)\lvec(7.5 -1.2)
    \move(7 -1)\lvec(7.5 -0.5)
    \move(7 0)\lvec(7.5 0)
     \move(7.5 1.2)\lvec(8 1.2)
      \move(7.5 0.8)\lvec(8 0.8)

          \move(6 -0.35)\textref h:C v:B \htext{$S$}

  \end{texdraw}
\end{center}
\caption{Cognitive graphs: a,b,c}
\end{figure}

 In principle, we can consider our Neural Pathway Model
an approximation of TNGS-model, see Edelman [37] on Theory of Neural Groups Selection. To combine TNGS
with our model, we should consider hierarchic chains that basis elements not singular neurons, but
some groups of neurons). So it will be Neuronal Group Pathway Model. In such a model
we shall use natural numbers $x_j=0,1,..., p-1,$ for coding of states of basic neuronal groups.

{\bf 6.5. Main cognitive features of the model:}

a). {\it Nonlocality (with respect to Euclidean geometry) of psychological functions.}

b). {\it Wholeness-integral evolution of body$\to$mind field $\varphi$.} \footnote{Compare
to  Bohmian-Hiley-Pilkk\"anen approach [11]-[13].}

c). {\it Sensation-thinking.} Since neural pathways go through the whole body, a part of a
pathway involved in a high level psychological function can be connected to e.g. skin-sensitivity.
Thus high order psychological functions also depend on various physiological stimuli.

d). {\it Interrelation of distinct psychological functions.} The central neuron $S$ of a cognitive graph
plays the role of the center of the system of coordinates. Other neurons can also be considered as
such centers. Therefore the same pathway contributes to distinct psychological functions. Thus evolution of various
psychological functions is simultaneous evolution based on huge interrelation of corresponding
cognitive graphs, see section 9.

e). {\it Emotion based reasoning.} Our pathway thinking model supports the
fundamental conjecture of A. Damasio, [34], that emotions play an important role in
the process of `reason-thinking'. Pathways going through centers   creating emotions can participate
in psychological function of a high order thinking process, e.g., proving mathematical theorems.
On the other hand, pathways going through reasoning-centers can go through some emotional center.
Thus reason participate in creation of emotions and vice versa.

\section{Dynamics in the mental space.}

{\bf 7.1. Mental state.} The body$\to$mind field $\varphi(z)$ describes important features of  functioning of the neural
system (in particular, its part located in brain). However, we will not be
concentrated on
the study of dynamics, e.g. (\ref{I2}), of the body$\to$mind field, since
$\varphi(z)$ describes the production of information by  the neural
system (in particular, its part located in brain) and not the
flow of mental information by itself. I would like to formulate this as

\medskip

{\bf{Mental Thesis:}}{\it{ The cognitive meaning of a mental point (with respect to a psychological function $f)$
does not depend on a neural pathway  that produces this mental point.}}

Mental activity is performed not in the pathway space,
cognitive graph, but in the mental space. Thus mental information does not remember its neurophysiological origin.
Mental Thesis
is supported (at least indirectly) by experimental evidences that
functions of some damaged parts of brain can be (in some cases)
taken by other parts of brain, see e.g. [34], [38], [39]. This thesis is also supported by neurophysiological evidences that
very different neural structures in brains of different species (e.g. fish and rat, [40]) can fulfill the particular
psychological function.\footnote{Of course, we should recall that by choosing the central neuron $S$ we chose the concrete
psychological function $f.$ Thus `the cognitive meaning' is related to this concrete psychological function. By choosing another
psychological function (a system of coordinates) we get another cognitive meaning.}

Mental Thesis might
be considered as a kind of anti-materialist thesis. We would not like to be at
such a position. We understand well that the relation between the brain (in fact, in our pathway model
- the whole body) and mind plays the crucial role in mental activity.
Mental Thesis should be considered as directed against the individual
deterministic relation between functioning physical neural pathways in the body and cognitive meaning of the corresponding
mental points. The absence of such individual determinism does not contradict to statistical determinism:

\medskip

{\bf{Thesis of Statistical Pathway Cognition.}}{\it{ The cognitive meaning of a mental point
(with respect to a psychological function $f)$  is
determined by the statistical probability of realization of this point in the ensemble
of pathways  $\Pi_f$ (the cognitive graph corresponding to $f).$}}

{\bf Remark 7.1.} (On the notion of probability) {\small During many years I studied
foundations of probability theory, see [41].
I know well the large diversity of viewpoints to the notion of probability. For me probability has nothing to
do with `potentiality', `measure of belief' and other perverse views. Probability is a statistical measure.
Let ${\cal E}$ be a large ensemble of, e.g., physical systems. Let these physical systems have some states.
These states are represented by points of the state space.
The probability of a state $x$ with respect to the ensemble ${\cal E}$ is given by the proportion:}
\[
{\bf{p}}(x)=\frac{\mbox{the number of systems having the state}\; x}{\mbox{the total number of
elements in the ensemble} }
\]
In our model {\it physical systems are centered neural pathways;} states are strings of firings throughout
pathways - mental points. An ensemble ${\cal E}$  is a cognitive graph; the state space is the
mental space. We suppose
that the cognitive meaning of a mental point $x\in M$ is determined by
the quantity
\[{\bf p}(x)=\frac{\mbox{the number of neural pathways that produce}\; x}{\mbox{the total number of neural
pathways in the cognitive graph} }\]
We call the ${\bf p}(x)$ (probabilistic) {\it{mental state}.} Here and in all following considerations
it is assumed that a psychological function $f$ is fixed. In fact, ${\bf p}(x)$ depends on $f:$
${\bf p}(x)={\bf p}_f(x).$

{\bf 7.2. Mental evolution.} Mental processes are  probability-evolution processes. These are  evolutions of mental states.
We have to find a mathematical model that would provide the adequate description of the evolution:
$t\to {\bf {p}}(t,x).$ Consider a discrete
dynamical system in the space of probability distributions:
\begin{equation}
\label{DEMS}
{\bf {p}}(t_{n+1}, x)=L {\bf {p}}(t_{n}, x),
\end{equation}
where $L$ is some operator in the space of probability distributions.
The generator of evolution $L$ may be linear, may be nonlinear. There must be performed experimental
studies to find the form of $L.$ Of course, $L$ depends essentially on
a psychological function $f$
and an individual.

Intuitively it is clear that there must be performed integration
over the whole mental space. The following evolution can be considered:
\[{\bf p}(t_{n+1}, x) =\int_{M} K(t_n, x, y){\bf p}(t_n, y)dy , \]
where $K(t,x,y)$ is a time-dependent kernel of evolution. For $p$-adic mental space $M={\bf Q}_p,$
$d y$ is the {\it Haar measure} -- an analogue of the ordinary linear
Lebesgue measure on the real line.

The mental state ${\bf p}(t,x)$ is nothing else than the probability distribution of the
body$\to$mind field $\varphi(z).$ We can consider the cognitive graph $\Pi_f$ as a probability
space $\Omega \equiv \Pi_f$ with the uniform probability measure
$$
{\bf P}(\omega)=\frac{1}{\mbox{number of elements in}\; \Omega}\;,
$$
for $\omega\in \Omega.$ Following to the probabilistic tradition we use the symbol $\omega$
to denote a point of the probability space. The map (body$\to$mind field) $\varphi:\Omega  \to M$ is a
random variable and the mental state
$$
{\bf p}(x)={\bf P}(\omega \in \Omega:\varphi(\omega)=x)
$$
gives
the probability that neural pathways in the cognitive graph represent the mental point $x.$
This is  intensivity of neural representation of $x$.\footnote{According to Kolmogorov's ideology [42]
(that gives the basis of the modern probability theory) the structure (e.g. topological)
of a probability space (in our case a cognitive graph $\Pi_f$) does not play any role in probabilistic
formalism. Probabilistic formalism depends only on the structure of the configuration space
in that random variables take values. In our case this is the $p$-adic mental space $M={\bf Q}_p.$}

Thus the dynamics of the mental state, $t\to {\bf p}(t,x),$ can be considered as the product of
the dynamics of the corresponding stochastic process $\varphi(t, \omega),$ the
process of body$\to$mind correspondence: ${\bf P}(t, U)={\bf P}(\omega\in \Omega:
\varphi(t, x) \in U),$ where $U$ is a domain in the ${\bf Q}_p$ (or in the cartesian
product of $p$-adic fields).
But(!) probability theory tells us that we could not reconstruct
the stochastic process $\phi(t, \omega)$ as a point wise map
in the unique way on the basis of corresponding
probability distributions. Two stochastically equivalent  body$\to$mind fields $\varphi_1(t,\omega)$
and $\varphi_2(t,\omega)$ produce the same dynamics of probabilistic mental state ${\bf p}(t,x),$
cf. section 2. This argument gives the strong support
to our Mental Thesis. In our model there exists the body$\to$mind field $\phi(t, \omega),$
but there is no mind$\to$body field.
This probabilistic consideration can be used as the strong argument supporting nonreductionism:
{\it Neural reduction of mental processes is impossible.}

\section{Diffusion model for dynamics of mental state}

We now consider continuous time evolution for
the simplest (from the mathematical viewpoint)  body$\to$mind fields -- mental diffusion.
\footnote{As we have already remarked, the ultrametic
topology on the mental configuration space and continuous real time dynamics are incompatible. However,
we do not have such a problem with probabilistic dynamics. Here both time and probability distributions
are real valued continuous parameters.}

The evolution of the mental state ${\bf p}(t,x)$ is described by the forward Kolmogorov equation
(\ref{CK}), where $L$ is a differential operator on the $p$-adic space generating diffusion.
Here the time $t \in [0, +\infty)$ is a real parameter and the mental point $x\in {\bf Q}_p$
is a $p$-adic parameter and the value ${\bf p}(t,x)\in [0,1]$ is also a real number.

{\bf Remark 4.1.} (Determinism or free will?)
{\small Our mental model combines  determinism and free will.
Mental determinism is a consequence of deterministic evolution equations for probability
distributions. However, this determinism
is determinism of probabilities. Thus probabilistic representation of the mental state gives
feeling of free will. We remark that the situation has some similarities with quantum theory, see [43]
and extended bibliography in this paper.

In the $p$-adic  mental space $M$ the simplest diffusion process is
Vladimirov-Volovich diffusion. This process was intensively studied in $p$-adic theoretical physics, [33].
The corresponding  evolution equation ($p$-adic heat equation) has the form:
\begin{equation}
\label{V}
\frac{\partial {\bf p}}{\partial t}(t,x)=-\frac{1}{2} D^2_x {\bf p}(t,x), {\bf p}(0,x)={\bf p}_0(x),
\end{equation}
where $D_x$ is  a kind of differential operator on the $p$-adic tree (Vladimirov's operator, [33]).
We remark that (in the opposition of the ordinary real derivative) Vladimirov's operator is nonlocal,
i.e., $ D_x {\bf p}(t,x)$ contains summation over all points of the mental space. This
feature of $p$-adic diffusion could be related to wholeness of mental processes.

We can easily find the fundamental solution $K(t,x).$ This is the solution for the initial mental
state
${\bf p}_0(x)=\delta(x).$ Here $\delta(x)$ is well known Dirac's $\delta$-function.
This is the probability distribution that is
concentrated in the fixed mental point.
By knowing the dynamics, $K(t,x), x\in M,$ of the mental
state starting with the `sharp mental state' $\delta(x)$
(all neural pathways of the whole mental graph produce the same mental point)\footnote{We understood that
such a mental state might not be approached in reality. So $K(t,x)$ is merely the ideal object used in the
mathematical model.}. We can find the dynamics of the mental state ${\bf p}(t,x)$ for any initial
distribution ${\bf p}_0(x):$
\[{\bf p}(t,x)=\int_{M} K(t,x-y) {\bf p}_0(y) d y .\]

A more realistic model of mental evolution is based on the $p$-adic heat equation
with a mental potential $V.$ Here $V(x,y)$ can be chosen as e.g.
$$
V(x,y)= \rho_p(x,y)^\alpha, \; \; \alpha\geq 0,
$$
where $\rho_p(x,y)$ is the $p$-adic distance between mental points.

\section{Discussion}

{\bf 9.1.  Can be consciousness be treated as a variable?}
The first part of the book [44] of B. J. Baars contains an interesting discussion on the possibility
to treat consciousness as a variable and the importance of such a treatment in cognitive science.
My attempt to quantify mental state was, in particular, motivated by this discussion. However,
we essentially modified  Baars' idea on mental-variable. As I understood, his
consciousness-variable would be a kind of Newtonian physical variable such as position, velocity,
force; see p. 11 [44] on similarity of consciousness-variable to variables in  Newtonian gravity. Our
mental-variable, mental state, is a probability distribution. Thus there are more
similarities with statistical mechanics and even with quantum mechanics. Such a variable
is not a local variable on brain. This is merely a {\it wholeness variable,} compare to [10]-[13].

Starting with a mental (probabilistic) state we can define a quantitative measure of consciousness,
a kind of consciousness-variable. Of course, we understood well that such a complex phenomenon as
consciousness could not be completely described by probability fields on mental space. Thus we just
propose a mathematical formalization of some features of consciousness. Maybe there can be found
many (may be infinitely many) other quantitative measures of various features of consciousness.

 First we discuss the connection between levels of neural activity
and levels of consciousness. The idea that there is the direct correspondence between
levels of  neural activity and  levels of consciousness is
the very common postulate of cognitive science. An extended discussion on
this problem can be found, for example, in [44], p.18-19. I disagree that there
is such a direct relation between levels of neural activity and  levels of consciousness.
For example, let us consider the extreme case in that all possible states of brain are activated.
Such a super-activation definitely would not imply a high level of consciousness.
It might be that not simply the level of neural activity determines the level of consciousness.

Our conjecture is that one of quantitative measures of consciousness is determined by the {\it variation} of a mental state
(compare to [10]). One of possible numerical measures of the level of consciousness
is {\it entropy} of a mental state. It is well known that (for the discrete probability
distribution) entropy approaches the maximal value for the uniform probability distribution.
By our interpretation this is the lowest level of consciousness (so it may be better to use
entropy with minus sign as a quantitative measure of consciousness).

Another possibility is to define a measure
of consciousness as the variation of a mental state
(with respect to the $p$-adic metric on the mental space). This is a natural quantitative measure
of some features of consciousness, since the topological structure of the mental space must be taken into account.
The $p$-adic variation can be defined by using Vladimirov's differential operator [33], $D_x,$ on the
$p$-adic tree:
\begin{equation}
\label{CV}
C({\bf p})=  \int_{M}\vert D_x {\bf p}(x)\vert^2 dx ,
\end{equation}
where ${\bf p}(x)$ is the mental state of a cognitive system. $C({\bf p})$ is always
nonnegative and it takes its minimal value, $C({\bf p})=0,$ for the uniform probability distribution.
At the moment we do not know anything about
mental states having the $C({\bf p})$ of the extremely high level, namely solutions of the  problem:
$C({\bf p})\to \rm{max}.$

{\bf 9.2. Are animals conscious?} Our model strongly supports the hypothesis that
animals are conscious (see e.g. [44] on the detailed neurophysiological and behavioral
analysis of this problem). If one of  quantitative measures of
 consciousness is really determined by the variation $C({\bf p})$ of the
mental state,  then animals are definitely able to produce such nontrivial variations by their
systems of neural pathways. Moreover, in our Neural Pathway Model pathways going throughout body play the
important role in the creation of consciousness. Thus the role of differences in brain structures
should not be overestimated, compare with [44]. On the other hand, animals have lower levels of
$C({\bf p}).$ We can quantify this problem in the following way. There exists a threshold $C_{\rm{human}}.$
Animals could not produce mental states ${\bf p}(x)$ such that $C({\bf p})$ is larger than
the human  threshold $C_{\rm{human}}.$ And human beings (at least most of them)
could produce (at least sometimes) mental states for that $C({\bf p})$ is larger than
this threshold.

{\bf 9.3. Blindsight.} Our model might be used to explain the mystery of blindsight and similar phenomena:
{\small ``The mystery of blindsight is not so much that unconscious visual knowledge remains.
...The greatest puzzle seems to be that information that is not even represented in area
$V1$ is lost to consciousness when $V1$ is damaged."} - [44]. However, in our Neural
Pathway Model by destruction of some individual neurons (e.g. in area $V1)$ we destroy (modify)
huge ensembles of neural pathways. Of course, by our model information was never
preserved in area $V1$ nor some other localized area. Information is
preserved by ensembles of pathways and they  are not located in some particular
domain of brain.

However, we also have to explain the unique function of area $V1$ in
creating of visual consciousness: {\small ``$V1$ is the only region whose loss abolishes our ability to consciously
see objects, events... But cells in $V1$ respond only to a sort of pointillist level of visual perception...
Thus it seems that area $V1$ is needed for such higher-level experiences, even though it does not
contain higher-level elements! It seems like a paradox."} - [44]. Yes, area $V1$ is the unique region that damage
destroy our ability to conscious visualization.  But we recall that our model is, in fact,
Centered Neural Pathway Model. The uniqueness of area $V1$ in  conscious visualization is determined by
the fact that central neurons of cognitive graphs corresponding to the psychological function of
conscious visualization are located in area $V1.$

We continue citation of Baars [44]: {\small ``Cells that recognize objects, shapes, and textures
appear only in much ``higher" regions of cortex, strung in a series of specialized regions along the bottom
of the temporal lobe."} Yes, these cells are centers of cognitive graphs corresponding to other
psychological functions, e.g. object recognition.

{\bf 9.4. Neural code and structure of mental space}. Suppose that the coding system of a cognitive
system  is based on a frequency code. There exists a time interval $\Delta$
(depending on a cognitive system and a psychological function) such that
a mental point produced by a centered neural pathway
is a sequence with coordinates given by numbers of oscillations for corresponding neurons during
the interval $\Delta.$ Thus in our model the problem of the neural code is closely related to the problem
of time-scaling in the neural system. For different $\Delta,$ we get different coding systems, and, consequently,
different structures of mental spaces. The corresponding natural number $p$ that determines the $p$-adic
structure on the mental space is defined as the maximal number of oscillations  that could be performed
by neurons (in the cognitive graph for some fixed psychological function) for the time interval $\Delta.$
The frequency coding is based on the 2-adic system induces the 2-adic mental space
that differs crucially from  the 5-adic (or 1997-adic) mental space induced by
the 5-adic (or 1997-adic) system. As it was demonstrated in [1],   by changing the $p$-adic
structure we change crucially dynamics. Hence the right choice of the time scaling parameter
$\Delta$  plays the important role in the creation of an adequate mathematical model for
functioning of a psychological function.

{\bf 9.5. Psychological time.} There might be some connection between the time scale parameter
$\Delta$ of neural coding and {\it psychological time.} There are strong experimental evidences,
see e.g. [45], that a moment in psychological time correlates with $\approx 100$ ms of physical time
for neural activity. In such a model the basic assumption is that the physical time required for
the transmission of information over synapses is somehow neglected in the psychological time.
In the model, the time ($\approx 100$ ms) required for the transmission of information
from retina to the inferiotemporal cortex (IT) through the primary visual cortex (V1) is mapped to a moment
of psychological time. It might be that by using $\Delta \approx 100$ ms we shall get the right
$p$-adic structure of the mental space.

Unfortunately, it seems that the situation is essentially more complicated. There are experimental evidences
that the temporal structure of neural functioning is not homogeneous. The time required for completion of color
information in V4 ($\approx 60$ ms) is shorter that the time for the completion of shape analysis in IT
($\approx 100$ ms). In particular it is predicted that there will be under certain conditions
a rivalry between color  and form perception. This rivalry in time is one of manifestations of complex
level temporal structure of brain. It can be shown that at any given moment in physical time, there
are neural activities in various brain regions that correspond to a range of moments in psychological time.
In turn, a moment in psychological time is subserved by neural activities in different brain regions at different
physical times.

Therefore it is natural to suppose that different psychological functions have different time scales
and, consequently, different mental spaces. Thus one psychological function is based on
the 2-adic mental space and another on the 5-adic (or 1999-adic) mental space. This is the very
delicate point and we shall try to clarify it.

There is the total space $\Pi$ of all neural pathways. The concrete psychological function
$f$ is based on some centered cognitive graph $\Pi_f$ (a subset of $\Pi$). There exists the fixed time scale
$\Delta= \Delta_f$ corresponding to this psychological function. Hence there exists the natural number
$p$ depending on $\Delta_f$ (and hence on the $f)$ determining the $p$-adic structure of the mental
space for the $f.$ Thus $p=p_f.$ On this $p_f$-adic mental space there is defined the mental state
${\bf p}_f(x)$ of the $f.$ In general another psychological function $g$ has its own time scale
$\Delta_g$ and corresponding $p_g.$ Its mental state ${\bf p}_g(x)$ is well defined on the $p_g$-adic space.
If $p_f$ is not equal to $p_g$ (e.g. $p_f=2$ and $p_g=1997),$
then  dynamics of mental states corresponding to psychological functions $f$ and $g$ differs
crucially -- even if evolutions are described by the same e.g. diffusion equation.

Finally, we remark that many psychological functions are strongly inter related on the neural
pathway level. Cognitive graphs $\Pi_f$ and $\Pi_g$ corresponding to
psychological functions $f$ and $g$ can have large intersection. In the extreme
case these graphs could even coincide: $\Pi_f= \Pi_g.$ But the use of different
time scales $\Delta_f \not= \Delta_g$ would produce totally different
evolutions for corresponding mental states.

{\bf 9.6. Discreteness of time.} Previous considerations demonstrated that the continuous
real time model for the evolution of the mental state  gives only rough approximation to the
really performed discrete time evolution. Moreover, discretization steps depend on
corresponding psychological functions.

Thus the evolution of the mental state ${\bf p}_f(t,x)$ of a psychological function
$f$ is described by discrete time dynamics, where
$t_{n+1}= t_n+ \Delta.$

{\bf 9.7. What is about $p$-adic structures of our psychological functions?} If we accept Edelman's
Neural Darwinism [37], then we have to consider the possibility that $p$-adic structures of our
psychological functions can depend on individuals. Thus, for an individual ${\cal I}$, the basis
of his/her mental space  for a psychological function $f$ depends both on
$f$ and ${\cal I}: \; p= p_{f, {\cal I}}.$ For two different individuals, for example,
Ivan and Andrei, the same psychological function $f$ can be based on two different
mental spaces, $M_{\rm{Ivan}}$  and
$M_{\rm{Andrei}}$, that are the $p_{f, \rm{Ivan}}$-adic tree and the $p_{f, \rm{Andrei}}$-adic tree,
respectively.

{\bf 9.8. Does consciousness benefit from long neural pathways?} Finally, we discuss one of the
greatest mysteries of neuroanatomy, see, for example,  [34], [37]. It seems
that in the process of neural
evolution cognitive systems tried to create for each psychological function
as long neural pathways as possible. This mystery might be explained on the basis of our neural
pathway coding model. If such a coding be really the case, then a cognitive system $\tau$ gets
great benefits by extending neural pathways for some psychological function as long as possible.
For example, let the neural code be based  on $p=5$ and let a psychological function $f$ be based on very short
pathways of the length $L=2.$ Then the corresponding mental space contains $N(5,2)= 2^5= 32$ points.
Let now $p=5$ and $L=10000.$ Then the corresponding mental space contains huge number of
points $N(5,10000)= 10^{20}$ points. On the latter (huge) mental space there can be realized mental states
having essentially more complex behavior (and, consequently, higher magnitude of consciousness).
This `mental space extending' argument can be used to explain spatial separation of various maps in brain,
see e.g. Edelman [37].

{\bf 9.9. Why  activity of ``far away" neurons  play important role?}
Consider a psychological function based on a cognitive graph with one central
neuron $S.$ Suppose that interaction between mental points (produced by the graph)
depends on the $p$-adic distance between these
points, e.g.
$$
V(x,y)= \rho_p (x, y)^\alpha, \; \alpha> 0.
$$
Then changes of states of input neurons that are located far away from the central neuron $S$ (on neural pathways
belonging to the cognitive graph) play the crucial role in variation of the
magnitude of the mental potential $V(x,y).$  If  states (e.g. rates of firing)
of initial input neurons are different, then $\rho_p (x, y)$ is very large, see  (\ref{PM1}).

\section{Postulates}

Our mathematical model of probabilistic thinking on $p$-adic mental spaces\footnote{ Such spaces  are produced by
cognitive graphs of hierarchic neural pathways.}
is based on following four postulates:

\medskip

{\bf 1. Pr} : {\it Mental states are determined by probability distributions on mental spaces.}

Evolution of mental states is described by classical (or may be even quantum)
evolution equations for probability distributions of random processes, e.g.
diffusion equations, on ultrametric $p$-adic mental spaces.

\medskip

{\bf 2. NeurPath}: {\it `Quant' of mental information (point of mental space) is given by the state of a hierarchic
neural pathway.}
\medskip

{\bf NeurGr}: {\it Each psychological function is based on a hierarchic graph of neural pathways,
cognitive graph.}

\medskip

{\bf 3. Ult}: {\it Mental topology is ultrametric.}

It is supposed that mental spaces (in the opposite to spaces used in physical
models) have ultrametric topology. The presence of such geometry is equivalent to a
treelike representation of mental space.

\medskip

{\bf 4. FrCod}: {\it Mental encoding of information is performed by accounting frequencies of
firings of neurons throughout hierarchic neural pathways.}

This encoding of information determines the natural number $p$ (depending on a psychological function and
a cognitive system).

The first postulate, {\bf Pr}, determines the (probabilistic) structure of high level information
processing in cognitive systems. Other postulates are related to  processing of cognitive information
on the primary level.
In fact, {\bf Pr} need not be rigidly connected with further postulates. There can be developed
other cognitive models of probabilistic thinking. In particular,  we need not base all models on the last postulate
{\bf FrCod}. There can be other models of  mental coding.

Finally, we briefly discuss relation of our Probabilistic Neural Pathway Model to some
traditional models of cognition. As was already remarked, we do not study
neural dynamics in the physical brain. Our model is purely information model. We study flows of
specially organized information. Postulates {\bf NeurPath} and {\bf FrCod}
provide connection with neurophysiology. However, we are not interested in investigation of functioning
of neural networks producing hierarchic strings of information, mental points. The only important thing
is the form of probability distribution (mental state) on the space of mental points. As it was
already underlined, different dynamical processes on neural level can produce
the same probability distribution.

In principle, there can exist some model,
for example,  {\it connectionist} (neural network) model, that would describe ``production'' of
information strings forming a mental state. However, such a generalized connectionist model should be
based on new paradigm: hierarchic neural pathway as basis processing unit (not single neuron!). We even can
not exclude the possibility
that such an ``underground model'' could be some {\it AI-model}. Our experience of computer simulations
shows that very complex random behaviour can be algorithmically simulated. But, in principle,
we need not presuppose
existence of any deterministic ``underground model".

I also would like to point to some connection with {\it distributed representation} models.
We recall that {\small A distributed representation is one in which
meaning is not captured by a single symbolic unit, but rather arises from the interaction of a set of
units, normally a network of some sort} . If we use just the first part of this definition (i.e., omit
relation to neural networks), then we could
consider Probabilistic Neural Pathway Model as a kind of model of distributed representation:
mental units (points) are unified through probability distribution -- mental state.

\medskip

I would like to thank all people (neurophysiologists, physicists, philosophers, psychologists, mathematicians,
parapsychologists) with whom I discussed (at various occasions) different aspects of my model.
I am especially grateful to  D. Amit, B. Hiley, S. Greenfield, G. Parisi, P. Pilkk\"anen,
A. Plotnitsky, G. Vitiello and all participants of seminars at University of Rome (``La Sapineza'',
Dept. of Psychology), Moscow State University (the Chair of High Neural Activity) and Institute of High Neural Activity
and Neurophysiology of Russian Academy of Science devoted to presentations and discussions of my $p$-adic
model.

\medskip

{\bf References}

1. A. Yu. Khrennikov,  {\it Non-Archimedean analysis: quantum paradoxes, dynamical
systems and biological models.} Kluwer, Dordrecht (1997).

2. A. Yu. Khrennikov, Human subconscious as the $p$-adic dynamical
system. {\it J. of Theor. Biology,} {\bf 193}, 179-196 (1998) .

3. A. Yu. Khrennikov,  $p$-adic dynamical systems: description of concurrent struggle in biological
population with limited growth. {\it Dokl. Akad. Nauk,} {\bf 361}, 752-754 (1998).

4.  S. Albeverio, A. Yu. Khrennikov, P. Kloeden, Memory retrieval  as
a $p$-adic dynamical system. {\it  Biosystems,} {\bf 49}, 105-115 (1999).

5. A. Yu. Khrennikov,  Description of the operation of the human subconscious by means
of $p$-adic dynamical systems. {\it Dokl. Akad. Nauk,} {\bf 365}, 458-460 (1999).

6. D. Dubischar D., V. M. Gundlach, O. Steinkamp,
A. Yu. Khrennikov, A $p$-adic model for the process of thinking
disturbed by physiological and information noise. {\it J. Theor. Biology,}
{\bf 197,} 451-467 (1999).

7. A. Yu.  Khrennikov,
$p$-adic discrete dynamical systems and collective behaviour of
information states in cognitive models. {\it Discrete Dynamics in Nature and Society,} {\bf 5,} 59-69 (2000).

8. S. Albeverio, A Yu. Khrennikov,
B. Tirozzi, $p-$adic Neural Networks, {\it Mathematical models and methods in
applied sciences}, {\bf 9},  1417-1437 (1999).

9. A. Yu. Khrennikov,  Classical and quantum mechanics on information spaces
with applications to cognitive, psychological,
social and anomalous phenomena. {\it  Found. Phys.} {\bf 29,}  1065-1098 (1999).

10. A. Yu. Khrennikov,  Classical and quantum mechanics on $p$-adic trees of ideas.
{\it BioSystems,}  {\bf 56}, 95-120  (2000).

11. D. Bohm,  and B. Hiley , {\it The undivided universe:
an ontological interpretation of quantum mechanics.}
Routledge and Kegan Paul,  London (1993).

12. B. Hiley, P. Pylkk\"anen,  Active information and cognitive science --
A reply to Kiesepp\"a. In: {\it Brain, mind and physics.}  Editors: Pylkk\"anen, P., Pylkk\"o, P., Hautam\"aki, A..
IOS Press, Amsterdam (1997).

13. B. Hiley, Non-commutavive geometry, the Bohm interpretation and the mind-matter relationship.
Proc. CASYS 2000, Liege, Belgium (2000).

14. R. Ashby, {\it Design of a brain.} Chapman-Hall, London (1952).

15. T. van Gelder, R. Port, It's about time: Overview of the dynamical approach to cognition.
in {\it Mind as motion: Explorations in the dynamics of cognition.}
Ed.: T. van Gelder, R. Port. MITP, Cambridge, Mass, pp. 1-43 (1995).

16. S. H. Strogatz, {\it Nonlinear dynamics and chaos with applications to physics,
biology, chemistry, and engineering.} Addison Wesley, Reading, Mass. (1994).

17. R. Port, The dynamical systems hypothesis in cognitive science.
{\it MacMillan encyclopedia of cognitive science.} To be published.

18. C. Eliasmith, The third contender: a critical examination of the dynamicist
theory of cognition. {\it Phil. Psychology,} {\bf 9(4)}, 441-463 (1996).

19. T. van Gelder, What might cognition be, if not computation?
{\it J. of Philosophy,} {\bf 91} 345-381 (1995).

20. A. Newell and H. Simon, Computer science and empirical inquiry.
{\it Communications of ACM,} pp. 113-126 (1975).

21. N. Chomsky, Formal properties of grammas. Ed.: R. D. Luce, R.R. Bush, E. Galanter.
{\it Handbook of mathematical psychology,} {\bf 2}, Wiley, New York, pp. 323-418 (1963).

22. P.S. Churchland and T. Sejnovski,
{\it The computational brain.} MITP, Cambridge (1992).

23. D. Amit, {\it Modeling Brain Function.} Cambridge
Univ. Press, Cambridge (1989).

24. J. J. Hopfield,{\it Neural networks and physical systems with
emergent collective computational abilities, Proc. Natl. Acad. Sci. USA},
{\bf 79},  1554-2558 (1982).

25. W. Bechtel, A. Abrahamsen, {\it Connectionism and the mind.}
Basil Blackwell, Cambridge (1991).

26. Aristotle, {\it Categories}, written 350 B.C.E., translated by E.M. Edghill,
http://classics.mit.edu/Browse/browse-Aristotle.html

27.  W. Schikhof , {\it Ultrametric calculus.} Cambridge Univ. Press,
Cambridge (1984)

28. A. J. Lemin, The  category of ultrametric spaces is isomorphic
to the category of complete, atomic, tree-like, and
real graduated lattices LAT. {\it Algebra universalis}, to be published.

29. M. Mezard, G. Parisi, M. Virasoro, {\it Spin-glass theory and beyond.}
World Sc., Singapore (1987).

30. G. Parisi, N. Sourlas, $p$-adic numbers and replica smmetry breaking.
{\it Europ. Phys. J.}, {\bf 14B}, 535-542 (2000).

31. V. A. Avetisov, A. H. Bikulov, S. V. Kozyrev,
Application of $p$-adic analysis to models of breaking
of replica symmetry. {\it J. Phys. A: Math. Gen.,} {\bf 32}, 8785-8791 (1999).

32.   A.  Escassut, {\it Analytic elements in $p$-adic
analysis.} World Scientific, Singapore (1995).

33. V. S. Vladimirov, I. V. Volovich, E. I.    and Zelenov,
{\it $p$-adic Analysis and  Mathematical Physics.} World Scientific Publ.,
Singapore (1994).

34. A. R. Damasio,
{\it Descartes' error: emotion, reason, and the human brain.} Anton Books, New York (1994).

35. F. C. Hoppensteadt,  {\it  An introduction to the mathematics of neurons:
modeling in the frequency domain.}  Cambridge Univ. Press,
New York (1997).

36. F. C. Hoppensteadt  and E. Izhikevich,  Canonical models in
mathematical neuroscience. {\it  Proc. of Int. Math. Congress}, Berlin,
{\bf 3}, 593-600 (1998).

37. G. M.  Edelman, {\it The remembered present: a biological theory of consciousness.}
New York, Basic Books (1989).

38.   H. Damasio and A. R. Damasio, {\it Lesion analysis in neuropsychology.}  Oxford Univ. Press, New-York  (1989).

39. Fuster, J. M. D. (1997)
The prefrontal cortex: anatomy, physiology, and neuropsychology of the frontal lobe.

40. A. Clark, {\it Psychological models and neural mechanisms. An
examination of
reductionism in psychology.} Clarendon Press. Oxford (1980).

41.  A. Yu. Khrennikov, {\it Interpretations of Probability.}
VSP Int. Sc. Publishers, Utrecht/Tokyo, 1999.

42.  A. N. Kolmogoroff, {\it Grundbegriffe der Wahrscheinlichkeitsrechnung.}
   Springer Verlag, Berlin (1933); reprinted:
   {\it Foundations of the Probability Theory}.
 Chelsea Publ. Comp., New York (1956).

43. A. Yu. Khrennikov, Quantum-like formalism for cognitive measurements.
Proc. Int. Conf. {\it Quantum Theory: Reconsideration of Foundations.}
Series {\it Math. Modeling in Phys., Engineering and Cognitive Sc.}, {\bf 2},
V\"axj\"o Univ. Press, V\"axj\"o (2002).

44. B. J. Baars,  {\it In the theater of consciousness. The workspace of mind.}
Oxford University Press, Oxford (1997).

45. K. Mori, On the relation between physical and psychological time.
Proc. Int. Conf. {\it Toward a
Science of Consciousness,} p. 102, Tucson, Arizona (2002).

\end{document}